# High performance Tunnel Field Effect Transistors based on in-plane transition metal dichalcogenide heterojunctions

Jean Choukroun[1], Marco Pala[1], Shiang Fang[2], Efthimios Kaxiras[2,3] and Philippe Dollfus[1]

[1] Centre for Nanoscience and Nanotechnology, CNRS, Univ. Paris Sud, Université Paris Saclay, Orsay, France
[2] Department of Physics, Harvard University, Cambridge, Massachusetts 02138, USA
[3] John A. Paulson School of Engineering and Applied Sciences, Harvard University, Cambridge, Massachusetts 02138, USA

E-mail: jean.choukroun@c2n.upsaclay.fr

## Abstract

In-plane heterojunction tunnel field effect transistors based on monolayer transition metal dichalcogenides are studied by means of self-consistent non-equilibrium Green's functions simulations and an atomistic tight-binding Hamiltonian. We start by comparing several heterojunctions before focusing on the most promising ones, i.e $WTe_2$-$MoS_2$ and $MoTe_2$-$MoS_2$. The scalability of those devices as a function of channel length is studied, and the influence of backgate voltages on device performance is analysed. Our results indicate that, by fine-tuning the design parameters, those devices can yield extremely low sub-threshold swings (< 5mV/decade) and $I_{ON}$/$I_{OFF}$ ratios higher than $10^8$ at a supply voltage of 0.3V, making them ideal for ultra-low power consumption.

Keywords: TFET, transition metal dichalcogenides, $MoS_2$, low SS, quantum simulation, NEGF, tight-binding

## 1. Introduction

Power consumption is one the main limiting factors of progress in computing technologies, and the scaling of the power supply is the most effective approach to improve energy efficiency, as a ten-fold reduction in $V_{DD}$ results in a hundred-fold save in dynamic power[1]. However, maintaining a high ON/OFF current ($I_{ON}$/$I_{OFF}$) with a lower power supply requires an extremely steep transition between the OFF and ON state of the device, which standard MOSFETs simply cannot provide due to their working mechanism.
Thanks to their ability to yield subthreshold swings (SS) below the thermionic limit of 60 mV/dec at room temperature that constrains MOSFETs[1,2], Tunnel Field Effect Transistors (TFETs) are recognized[3] to be one of the most promising avenues for the aforementioned scaling of the power supply ($V_{DD}$). However, since TFETs rely on a band-to-band tunneling (BTBT) mechanism, the current they provide in the ON-state is often several orders of magnitude lower than that of MOSFETs -depending on the length of the depletion region to be tunneled- which severely constrains the possible applications[4].
As will be detailed in Sec.3, thanks to the bandstructure properties of the heterostructures investigated, the TFETs presented in this study do not suffer from this drawback and the $I_{ON}$/$I_{OFF}$ ratios they present are actually higher than that of most MOSFETs. Encouraging experimental results have been reported in the case of Si and III-V semi-conductor based TFETs[5–8], but the use of bulk materials entails a high concentration of traps and a high roughness at the interface, as well as dangling bonds, which all contribute to increasing the SS[9-11] and therefore decreasing the device perfomance. Moreover, in bulk materials, the quantum confinement arising from the nanoscale of the device widens the band gaps and prevents the formation of a truly broken band-gap heterostructure[12]. It is not the case for heterojunctions of 2D materials where, as described later, strain effects can actually induce a broken gap, which is most convenient for TFET performance, in particular in terms of $I_{ON}$.

Monolayer-based TFETs can be split into two categories : van der Waals TFETs, in which the monolayers are stacked vertically[13,14], and conventional "lateral" TFETs[15,16], in which the monolayers occupy the same plane. Several in-plane 2D heterostructures have already been experimentally realized: from graphene-hBN [17-20] to graphene-monolayer transition metal dichalcogenide (TMD)[18,21], to TMD-TMD[18,22,23], and the growth and deposition techniques related to 2D materials are rapidly expanding and becoming more versatile. Because of the aforementioned inherent advantages they hold when compared to bulk materials and of the recent advances in the techniques related to their experimental deposition, we elected to use in-plane 2D material heterostructures in the TFETs investigated. The materials used are monolayer transition metal dichalcogenides: semi-conductors with band gaps ranging from ~1 to 2 eV. Those materials as well as their reaction to strain were modelled via the tight-binding (TB) model detailed in Sec. 2.

In this article, we present atomistic quantum simulations of electronic transport in in-plane heterojunction TFETs based on TMDs as well as a pure $WTe_2$ TFET to be used as reference. Their transport characteristics (SS, $I_{ON}/I_{OFF}$) are then compared in order to select the most promising heterojunctions, which will be studied further. Namely, the influence of design parameters (backgate voltages, channel length) on their performance will be evaluated.

The paper is organized as follows: in Sec. 2, the TB model and materials used are introduced, and the device structure and simulation methodology are described; in Sec. 3, the simulation results for the pure $WTe_2$ TFET and all heterojunction TFETs are compared, while the most promising of these devices is studied further in Sec. 4. Finally, conclusions and future works are addressed in Sec. 5.

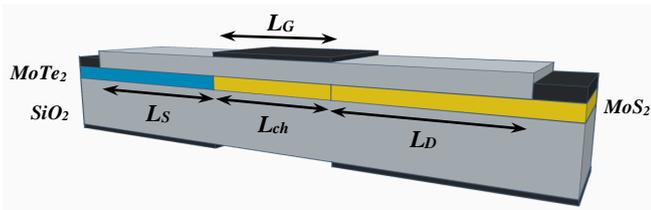

*Fig. 1. 3D sketch describing the structure of the studied TFETs (the $MoTe_2$-$MoS_2$ heterojunction is shown in this case). $L_G$, $L_S$, $L_{ch}$, and $L_D$ are the gate, source, channel and drain lengths, respectively.*

| MX$_2$ | MoS$_2$ | WS$_2$ | MoSe$_2$ | WSe$_2$ | MoTe$_2$ | WTe$_2$ |
|---|---|---|---|---|---|---|
| *a* (Å) | 3.18 | 3.18 | 3.32 | 3.32 | 3.55 | 3.55 |
| $E_{gap}$ (*eV*) | 1.79 | 1.95 | 1.55 | 1.65 | 1.25 | 1.23 |

*Table 1. Physical parameters of considered TMDs in their pristine form. The band gaps are calculated from the TB Hamiltonian, and the lattice parameters are taken from [25,26].*

## 2. Device description and simulation methodology

### 2.1 Device description

All of the modelled TFETs share the same structure, which is shown in Fig.1 ($WTe_2$-$MoS_2$ heterojunction in this case). It features a monolayer TMD source, channel and drain, a 3.35 nm thick $SiO_2$ buried oxide and a high-$\kappa$ top-gate oxide of equivalent oxide thickness $t_{oxe} = 0.44$ nm. In the case of the heterostructures, a first TMD acts as the source, while a second, different TMD is used in the channel and drain regions; the interface therefore lies between the source and the channel. Thanks to the 2D nature of the device, we can use backgates instead of chemical doping in order to control charge densities in the contacts, which allows for much more precise control over the behavior of the device and, contrary to chemical doping, does not introduce impurities in the material. Current flow through the device is controlled via a top-gate of length equal to the channel region.

### 2.2 Material modelling methodology

In this work, we consider five monolayer TMDs : $MoS_2$, $MoSe_2$, $WSe_2$, $MoTe_2$ and $WTe_2$, modelled using an 11-band tight-binding model presented in [24], in which the effect of strain on the electronic properties is taken into account. It is worth noting that while $WTe_2$ and $MoTe_2$ were not included in those studies, the same work has since been done with those materials in order to obtain the necessary TB parameters. In this TB model, all the relevant orbitals near the Fermi level -i.e the *p* orbitals for the chalcogen atoms and the *d* orbitals for metal atoms- are taken into account. The resulting TMDs are semi-conductors with direct band gaps ranging from 1.2 to 1.95 eV (see Fig. 2) located at the K-point of the Brillouin zone. Some relevant information regarding the pristine form of those materials can be found in Table 1 (the reported lattice parameters were taken from[25,26]). Most ab-initio studies[25-27] report lower band gaps for those materials because DFT notoriously underestimates band gaps[28]; as mentioned in the original article presenting the model, Green-Wannier calculations were performed to increase its accuracy, which explains the higher than average band gaps.

In the case of heterojunction TFETs, some strain has to be applied to the materials in order to obtain lattice matching at the interface. Our TB model takes strain into account (see Appendix 1 for details), and therefore allows us to apply the necessary stress to the considered material and compute the resulting electronic properties.

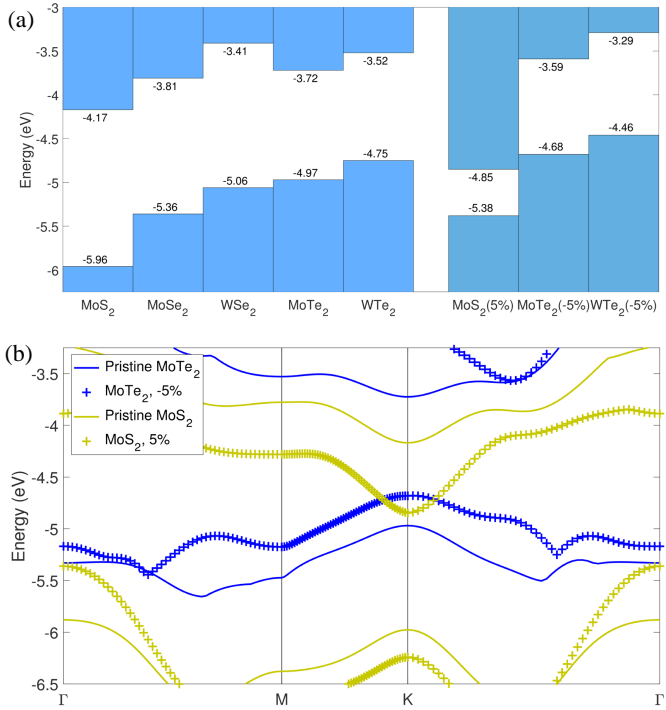

*Fig. 2. (a)Band alignment of the five considered TMDs, both in their pristine form (left) and when the strain needed in the heterostructures is applied (right). The vacuum level is set to 0 eV. It is worth noting that under these stresses, the band gaps become indirect in the considered materials. (b)Highest VB and lowest CB in pristine form (line) and under strain (crosses) for MoTe$_2$ (blue) and MoS$_2$ (gold). When the strains are applied, the top of the MoTe$_2$ VB is higher than the bottom of the MoS$_2$ CB, which is ideal for TFET operation.*

In the case of MoSe$_2$-MoS$_2$ and WSe$_2$-MoS$_2$ TFETs, a 4.26% tensile strain has to be applied to MoS$_2$ in order to reach a lattice parameter of 3.32 Å (the same as the source material); in the case of the WTe$_2$-MoS$_2$ and MoTe$_2$-MoS$_2$ TFETs, a 5% tensile strain was applied onto MoS$_2$, while a 5% compressive strain was applied to the source material, in order to reach a lattice parameter of 3.35 Å along the whole device. In similar systems (like WSe$_2$-MoS$_2$), it has been shown that the strain at the interface is relaxed over several tens of nanometers[29]. The lead-to-lead distance in the devices investigated being shorter than the relaxation length, we have assumed that the relaxation can be neglected, which makes the atomistic simulation of the full device computationally tractable.

The effect of strain on the electronic properties of the considered TMDs (namely bad gap and alignment) is shown in Fig.2 (a), and more specifically on the band structure in the case of MoS$_2$ and MoTe$_2$ in Fig.2 (b). We can see that, in the case of the WTe$_2$-MoS$_2$ and MoTe$_2$-MoS$_2$ devices, the top of the source valence band (VB) is actually located at a higher energy than the bottom of the channel conduction band (CB) when we apply the necessary stress to reach lattice matching. Because TFETs rely on band-to-band tunnelling (BTBT), the alignment of the valence and conduction bands between the source and the channel region is paramount: it dictates the length of the depletion region the carriers will have to tunnel through to reach the channel, and therefore severely impacts the performance of the device (both SS and ON current). By lowering its CB by approximately 70 meV, applying the aforementioned 5% tensile strain on MoS$_2$ is highly beneficial to the devices investigated. What is more, the 5% compressive strain on WTe$_2$ and MoTe$_2$ raises their VB by approximately 30 meV, so much so that they actually stand higher than the bottom of the MoS$_2$ CB; this configuration is known as a "broken gap". As will be shown in Sec.3, this is hugely beneficial to the device performance and is ideal for TFET operation, making the depletion region almost non-existent. It is worth noting that spin-orbit coupling is not included in this work, but is expected to raise the XTe$_2$ VB [30], increasing the overlap between the source VB and channel CB, and therefore benefiting the performance of the device.

### 2.3. Hamiltonian creation

As mentioned before, the TB model considers the $p_x$, $p_y$ and $p_z$ orbitals of the chalcogen atoms, and the $d_{xy}$, $d_{xz}$, $d_{yz}$, $d_{z^2-r^2}$ and $d_{x^2-y^2}$ of the metal atoms. As the unit cell ($MX_2$, represented in yellow in Fig.3) is composed of one metal atom and two chalcogen atoms, the initial basis is an 11x11 matrix (5 *d* orbitals + 2×3 *p* orbitals). In order to use the NEGF method, we need to describe the device as "layers", repeating along the transport direction, which forces us to use a bigger unit cell than the one used in the original TB model. This new unit cell is represented in red in Fig.3. It contains 2 metal atoms and 4 chalcogen atoms, and therefore leads to a 22×22 basis. Because of this change of unit cell, and of the way the model was introduced in the original article[24,30], some adaptation work was required to create matrices describing the coupling between different orbitals based solely on the positions of the atoms they are associated with. From those matrices, 22×22 matrices describing the Hamiltonian of a single unit cell ($H_{n,m}$ in Fig.3) and the coupling between this unit cell and an adjacent cell ($T_{i,j}$ in Fig.3) are deduced. We only represent the coupling for half of the adjacent cells in order to preserve readability, but the coupling are symmetrical with respect to the original ($n,m$) unit cell, so that $T_{n-1,m-1} = T_{n+1,m+1}^\dagger$ for instance. The matrices describing the orbital couplings and the creation of the 22×22 Hamiltonians are given in Appendix A. At a given $[k_x,k_y]$ wave vector, the "layer" Hamiltonians can be calculated as follows:

$$H_n(k_y) = H_{n,m} + T_{n,m+1} \cdot e^{i.k_y.a_y} + T_{n,m-1} \cdot e^{-i.k_y.a_y} \quad (1)$$

$$T_{n+1}(k_y) = T_{n+1,m} + T_{n+1,m-1} \cdot e^{-i.k_y.a_y} + T_{n+1,m+1} \cdot e^{i.k_y.a_y} \quad (2)$$

$$T_{n-1}(k_y) = T_{n+1}^\dagger(k_y) \quad (3)$$

From those Hamiltonians, the total Hamiltonian of the device is calculated as

$$H_{tot}(k_x, k_y) = H_n(k_y) + T_{n+1}(k_y) \cdot e^{i.k_x.a_x} + T_{n-1}(k_y) \cdot e^{-i.k_x.a_x} \quad (4)$$

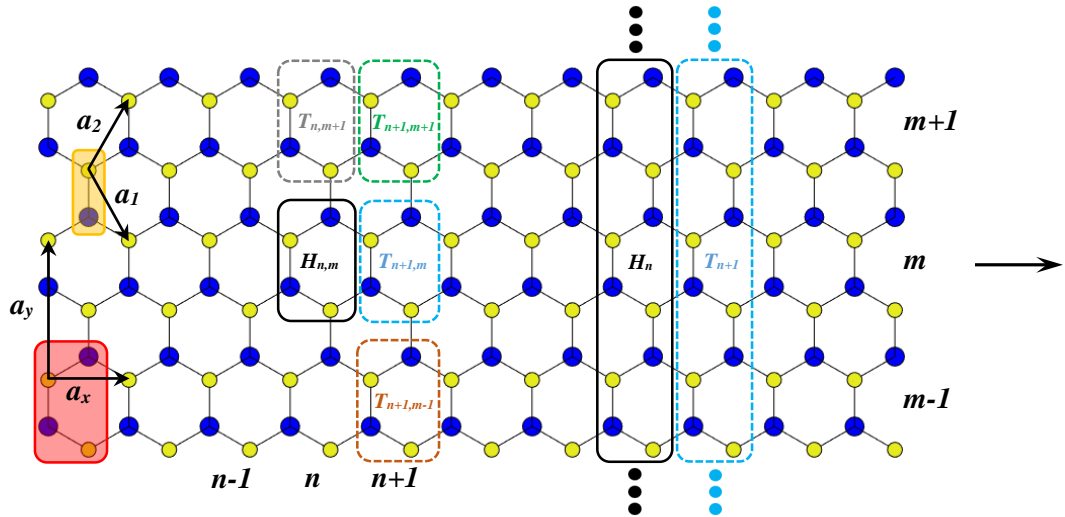

*Fig. 3. Atomic arrangement of TMDs. The yellow area represents the unit cell considered by the original TB model, and the red area represents the unit cell considered in our work. $H_{n,m}$ and $H_n$ are the hamiltonians for a unit cell and single "layer" of the material, respectively. Finally, $T_{i,j}$ and $T_i$ represent the coupling of the (i,j) cell with the (n,m) cell, and of the (i) layer with the (n) layer, respectively. Transport direction is indicated by the arrow.*

This total Hamiltonian describes the 2D infinite TMD layer at a given wave vector. In the case of an in-plane heterojunction, the appropriate orbital couplings are used in each material, and the coupling at the interface is calculated as the average of the coupling parameters of the materials on either side of the interface. As an example, the *(n+1,m)* coupling at the interface is calculated as $T_{n+1,m}^{A/B} = (T_{n+1,m}^{A} + T_{n+1,m}^{B})/2$ where A(B) is the TMD on the left (right) of the interface.

*2.4. Quantum simulation method*

In this work, we use the non-equilibrium Green's function (NEGF)[29,30] method self-consistently coupled with 3D Poisson equation to compute ballistic electronic transport through the simulated devices. Once the aforementioned tight-binding Hamiltonians have been generated, they can be used to calculate the device's Green function, from which we can calculate physical quantities such as current, charge, local density of states (LDOS) …

However, we use the Sancho-Rubio method[32] to calculate only the main diagonal and first sub-diagonal elements of the Green's function matrix, which are the only ones needed to obtain the physical quantities. This technique allows for an important reduction of the computational cost of those calculations, and is routinely used to simulate TFETs[31,32]. As mentioned before, this NEGF method is self-consistently coupled with the solving of the 3D Poisson equation: an initial guess of the potential profile is used to calculate the device Hamiltonian, from which we can obtain the device's Green function. From this matrix, we can calculate the charge densities in the device, from which is then deduced an updated potential profile to be used as input for the calculation of the updated device's Green function. This loop is repeated until coherence is reached. Mean-free paths around 20 nm have been reported for $MoS_2$[35], so, in the case of short devices, the ballistic approximation used here is expected to yield results comparable to those that would be obtained by including phonon scattering. Although the deformation potentials reported for TMDs are relatively small[36–38], phonon scattering will undoubtedly slightly impact performance in the case of devices of length exceeding 20 nm, by increasing SS -due to a widening of the density of states-, and decreasing ON current.

## 3. Results and discussion

In the following discussion of the results we will often refer to a specific metric to describe the performance of the investigated TFETs: their sub-threshold swing (SS). This metric is expressed in mV/dec, and describes the increase in gate voltage needed to increase the current tenfold, which is why the lower the SS, the steeper the slope. The SS of a logic device therefore relates to the steepness of the slope of the $I_{DS} - V_G$ characteristic at low $V_G$. Due to their working mechanism and the Fermi-Dirac distribution they are bound to, MOSFETs physically cannot provide sub-threshold swings below 60 mV/dec. Because TFETs rely on BTBT, they have no theoretical limit for SS and can approach the behavior of an ideal switch: to be in a fully OFF state at a given $V_G$, and in a fully ON state at an infinitesimally higher $V_G + \delta V_G$. This would allow for extremely fast and easy switching, requiring a minimal amount of energy. In this work, SS is calculated as the average swing between $I_{DS} = 10^{-5}$ and $10^{-2}$ µA/µm. The ON current, $I_{ON}$, is defined as the current at a gate voltage $V_{ON} = V_{OFF} + V_{DS}$, where $V_{OFF}$ is the gate voltage at which the value for the selected OFF current (in our case $10^{-5}$ µA/µm) is reached, and $V_{DS}$ is the drain bias applied to the device. The goal of the studied TFETs is therefore to provide the lowest SS and highest $I_{ON}/I_{OFF}$ ratio possible.

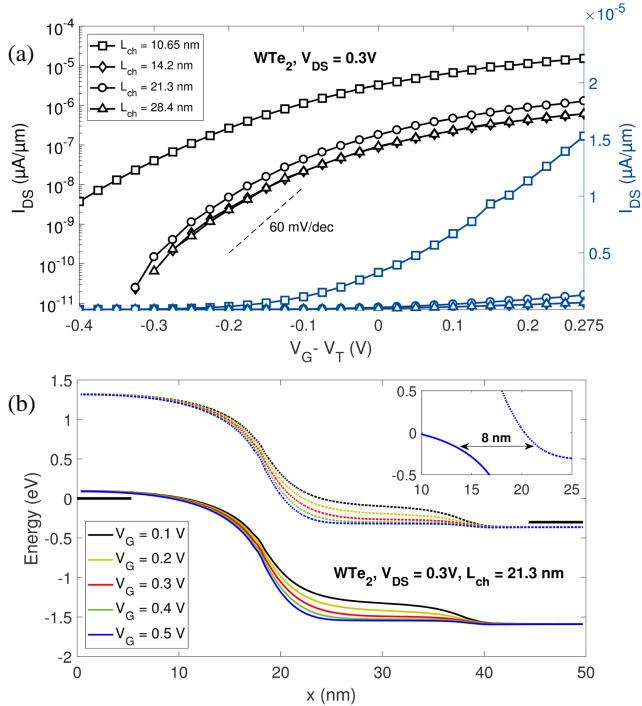

*Figure 4.(a)$I_{DS} - V_G$ characteristics of the pure WTe2 device (b)Highest CB and lowest VB for $V_G$ ranging from 0.1 to 0.5V; inset: zoom on the interface for $V_G = 0.5V$. Both figures were obtained at $V_{DS} = 0.3V$, $V_{BG-S} = -1.35V$, and $V_{BG-D} = -0.75V$.*

### 3.1. Pure WTe$_2$ TFET

Before studying in-plane TMD heterojunctions, we simulated a pure WTe$_2$ TFET so that it can be used as a point of reference and as comparison for the heterojunction based devices. We elected to use WTe$_2$ because it has the lowest band gap out of the five considered TMDs (1.23 eV in the TB model used), and should therefore have the shortest depletion region in the ON state and, subsequently, should present the highest performances. The WTe$_2$ TFET was studied at backgate voltages of $V_{BG-S} = -1.25V$, $V_{BG-D} = -0.75V$, at a supply voltage of $V_{DS} = 0.3V$, and has a 17.75 nm long source, a 10.65 nm long drain, and a channel length ranging from 10 to 29 nm. The corresponding $I_{DS} - V_G$ characteristics are shown in Fig. 4 (a), in which the threshold voltage ($V_T$) was obtained via the linear approximation of the ON state current. These characteristics highlight the poor performance of this pure WTe$_2$ device: the SS is higher than 60 mV/dec for all channel lengths considered, and low ON currents reaching only $10^{-5}$ µA/µm for the 10 nm channel, and $10^{-7}$ µA/µm for the longer ones. Fig. 4.(b), which shows the highest VB and lowest CB in this device for several gate voltages at $L_{ch} = 21.3$ nm, highlights the origin of these poor performances. Even at a high gate voltage of $V_G = 0.5V$, the depletion region the carriers have to tunnel through is approximately 8 nm long, which is too high for any significant current to take place. While it is the lowest band gap out of the five TMDs, the 1.23 eV gap of WTe$_2$ is too high and severely hampers the prospects of a pure WTe$_2$ device.

It is worth noting that several other works on similar devices have reported better performances [16,39,40], which can be attributed to the fact that, as mentioned in the presentation of the TB model, the 1.23 eV band gap calculated in this work is higher than routinely obtained DFT values, which usually underestimate the actual band gap[28].

### 3.2. Comparison of all heterojunction TFETs

The in-plane heterojunctions investigated are MoSe$_2$-MoS$_2$, WSe$_2$-MoS$_2$, MoTe$_2$-MoS$_2$ and WTe$_2$-MoS$_2$. As a reminder, MoS$_2$ is under a 4.2% tensile strain in the XSe$_2$-MoS$_2$ devices to reach $a_{Se} = 3.32$ Å; in the XTe$_2$-MoS$_2$ devices, MoS$_2$ is under a 5.3% tensile strain, while WTe$_2$ and MoTe$_2$ are under a 5.6% compressive stress, in order to reach a common lattice parameter of $a_{Te} = 3.35$ Å.

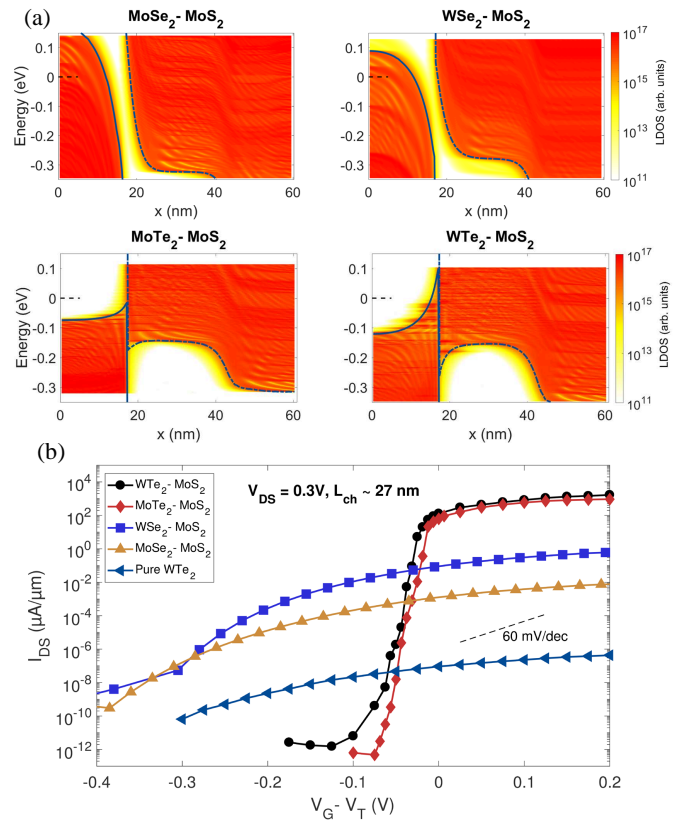

*Fig. 5 (a)LDOS at $V_G = 0.4$ V for all four heterojunction devices. (b)$I_{DS} - V_G$ characteristics for the pure WTe$_2$ TFET and all four heterojunction TFETs at $V_{DS} = 0.3V$, $L_S = L_D = 17$ nm and $L_{ch} = 27nm$.*

| Source MX$_2$ | MoSe$_2$ | WSe$_2$ | MoTe$_2$ | WTe$_2$ |
|---|---|---|---|---|
| Band offset (eV) | 0.65 | 0.35 | -0.17 | -0.39 |
| $L_{depletion}$ (nm) | 5.3 | 4.6 | ~0 | ~0 |
| $I_{ON}/I_{OFF}$ | 9×10$^2$ | 2×10$^4$ | 10$^8$ | 2×10$^8$ |
| SS (mV/dec) | 150 | 50 | <5 | <5 |

*Table 2. Band offsets (channel CB minimum – source VB maximum) and transport properties of the TFETs, at $V_{DS} = 0.3V$. $L_{depletion}$ was calculated at $V_G = 0.4V$.*

For the sake of clarity, the lengths mentioned in this part are rounded off, so that they apply to all devices. Fig. 4 (a) shows the local density of states, as well as the highest VB and lowest CB for all devices in the ON state, at a gate voltage of $V_G = 0.4V$, a supply voltage of $V_{DS} = 0.3V$, and lengths of $L_S = L_D = 17$ nm and $L_{ch} = 27$ nm. These LDOS figures give a clear picture of the depletion region the carriers have to tunnel through in the ON state, and highlights the benefits of the "broken gap" configuration found in the MoTe$_2$-MoS$_2$ and the WTe$_2$-MoS$_2$ devices: the depletion region is almost non-existent in those devices, which explains their outstanding performance, shown in Fig. 5 (b). This figure represents the $I_{Ds} - V_G$ characteristics of the pure WTe$_2$ and all heterojunction devices: the direct correlation between the length of the depletion region and the performance of the device is obvious; the XSe$_2$-MoS$_2$ devices show very low ON current and very high SS, due to their approximately 5 nm long depletion region. Out of those two, WSe$_2$-MoS$_2$ has the best performance, with a steeper slope in the OFF regime, and an ON current roughly 100 times higher than that of MoSe$_2$-MoS$_2$ TFET ; with that said, its $I_{ON}/I_{OFF}$ ratio ($2 \times 10^4$) is too low to realistically envision logic applications.

On the other hand, it is obvious from those characteristics that the XTe$_2$-MoS$_2$ devices are far more promising. They showcase an extremely steep slope in the OFF regime which leads to a $< 5$ mV/dec SS in both devices, and very high ON currents of roughly $10^3$ μA/μm in the case of MoTe$_2$-MoS$_2$, and $2 \times 10^3$ μA/μm in the case of WTe$_2$-MoS$_2$. The band offsets, depletion region lengths, $I_{ON}/I_{OFF}$ ratios and SS for all four heterojunction devices are summarized in Table 2.

Due to their show extremely promising performance (very low SS and high $I_{ON}/I_{OFF}$ ratio), the MoTe$_2$-MoS$_2$ and WTe$_2$-MoS$_2$ devices need to be studied more fully. For the MoTe$_2$-MoS$_2$ system especially, we will investigate the influence of design parameters such as channel length and backgate voltages on the device performance. The study of the WTe$_2$-MoS$_2$ device leading to very similar results, it is mentioned but not fully detailed here.

### 3.3. Study of the MoTe$_2$-MoS$_2$ heterojunction TFET

We start by studying the scaling of this device with respect to channel length, and its influence on transport properties. $I_{DS} - V_G$ characteristics for channel lengths ranging from 10 to 27 nm are shown in Fig. 5 (a), while Fig. 5 (b) highlights the impact of channel length on SS and ON current in this device. The current characteristics show the high impact of channel length on the steepness of the slope in the OFF regime: due to increased electrostatic integrity in longer channels, SS decreases from 60 mV/dec in the case of a short 7.4 nm channel to approximately 3 mV/dec when channel length exceeds 20 nm. The current behaves similarly at high gate voltages no matter the channel length, contrary to $V_{OFF}$ - the gate voltage at which $I_D = I_{OFF} = 10^{-5}$ μA/μm- , which is highly impacted. Therefore the ON current, calculated at $V_{ON} = V_{OFF} + V_{DS}$, increases with channel length until reaching a plateau around $10^3$ μA/μm for channel lengths exceeding 17 nm. LDOS and current densities for the 20.1 nm channel device are shown in Fig. 5 (c), and highlight the extremely low SS of this device. At $V_G = 0.375V$, the device is in a fully OFF state, as evidenced by the LDOS "gap" at the interface between $-0.04$ and $-0.06$ eV, and by a current density 8 orders of magnitude lower than in the ON state. At $V_G = 0.45V$ however, the device is in a fully ON state and current flows freely from the source to the drain, as represented in the current density figure.

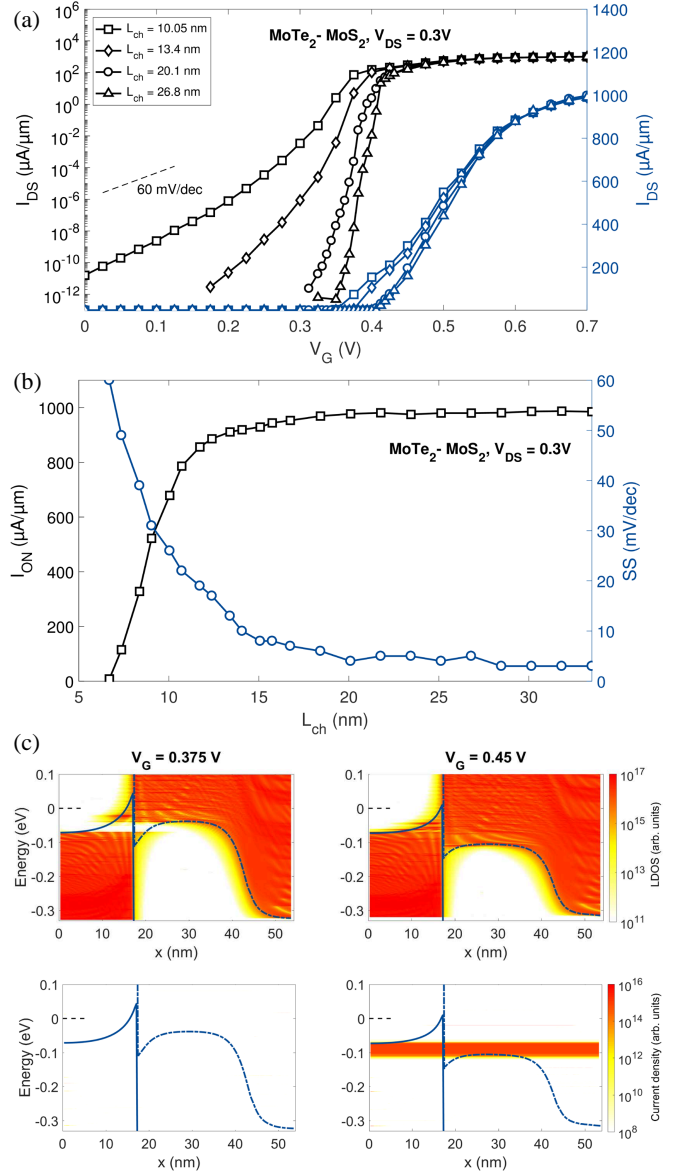

*Fig. 5. (a)$I_{DS} - V_G$ characteristics for the MoTe$_2$-MoS$_2$ TFET for several channel lengths (b)Influence of channel length on SS and ON current in this TFET. (c)LDOS and current density in the OFF (left) and ON (right) state for the TFET at $L_{ch} = 20.1$ nm; the highest VB (full lines)and lowest CB (dashed lines) along the device are also shown. In all figures, $V_{DS} = 0.3V$, $V_{BG-S} = 0.58V$ and $V_{BG-D} = 0.8V$.*

As mentioned when we described the structure of the device investigated, we elected to use backgates to electrostatically control charge densities in the contacts instead of chemical doping. Those backgates therefore directly control the energy states in the contacts, and are expected to be an important tuning parameter in this device. We will now study the influence of the source and drain backgate voltages on the performance of the MoTe$_2$-MoS$_2$ device with $L_{ch} = 13.4$ nm and $V_{DS} = 0.3$V. Fig. 6 (a) shows the current characteristics for this device at $V_{BG-D} = 0.8V$ and several source backgate voltages ranging from 0.4V to 0.58V, and the inset highlights the influence of $V_{BG-S}$ on the device's ON current.

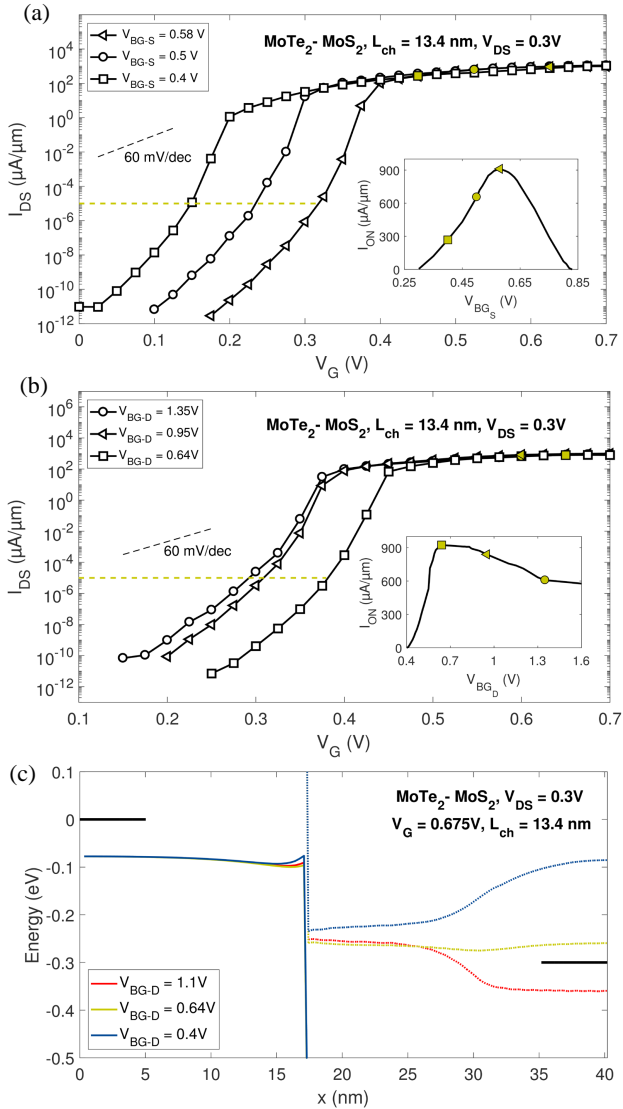

As we can see from the main figure, while $V_{BG-S}$ has no impact on SS, it highly controls the threshold voltage and therefore $V_{ON}$. The inset clearly shows a somewhat parabolic influence of source backgate voltage on ON current, with an optimal voltage range from approximately 0.55V to 0.6V. A similar study about the influence of drain backgate voltage was performed on the same device at a fixed $V_{BG-S} = 0.58$V. As is shown in Fig. 7(b), $V_{BG-D}$ also has a strong impact on the current characteristics of this device: SS decreases as we lower the drain backgate voltage, while the ON current increases to reach a maximum of approximately $9 \times 10^2$ µA/µm around $V_{BG-D} = 0.65$V. However, by decreasing the drain backgate voltage further, the CB in the drain is pulled towards higher energies, reducing the width of the tunneling window and therefore the ON current. As is shown in Fig. 7 (c), the drain CB actually rests higher than the source VB at $V_{BG-D} = 0.4$V, which explains the absence of current. The optimal drain backgate voltage therefore depends on the applied source backgate voltage.

We can conclude from this study that the optimal voltages to apply to the backgates to operate this in-plane MoTe$_2$-MoS$_2$ TFET at $V_{DS} = 0.3$V are $V_{BG-S} = 0.58$V and $V_{BG-D} = 0.64$V. Fig. 7 shows the $I_{DS} - V_{DS}$ characteristics of the 13.4 nm channel device at several gate voltages ranging from $V_G = 0.35$V to $V_G = 0.6$V. The current increases linearly with the applied gate voltage, and current saturation is reached around $V_{DS} = 0.2$V; this indicates that the device can operate at its full capacity even at low drain biases, which makes its use for ultra-low power operation even greater.

As presented, we were able to determine the optimal design parameters for this TFET in order to maximize its performance. By using a channel length of at least 20 nm and the aforementioned optimal backgate voltages, this device can yield a sub-threshold swing below 5 mV/dec and an $I_{ON}/I_{OFF}$ ratio of $10^8$. Those performances are far greater than those reported in other 2D material heterojunction based TFETs[5,12,16,39].

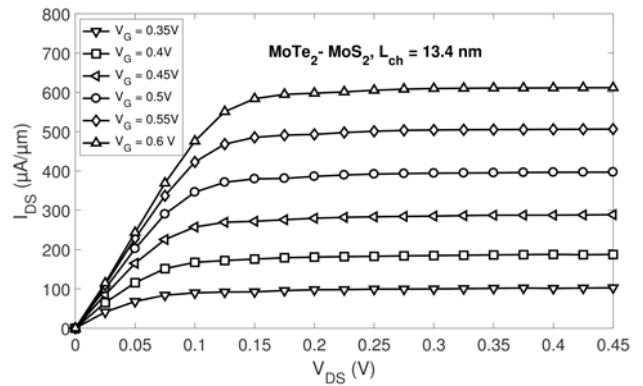

*Fig. 6. (a)$I_{DS} - V_G$ characteristics for the MoTe$_2$-MoS$_2$ TFET for several source backgate voltages at $V_{BG-D} = 0.8V$. Inset: influence of source backgate voltage on $I_{ON}$. (b)$I_{DS} - V_G$ characteristics for the MoTe$_2$-MoS$_2$ TFET for several drain backgate voltages at $V_{BG-S} = 0.58V$. Inset: influence of drain backgate voltage on $I_{ON}$. In both (a) and (b) figures, the yellow dashed line indicates $I_{OFF} = 10^{-5} µA/µm$, and the yellow markers indicate the ON current calculated at $V_{ON} = V_{OFF} + V_{DS}$. (c)Highest VB (full lines) and lowest CB (dashed lines) for the device shown in (b), at several $V_{BG-D}$, and $V_G = 0.675V$.*

*Fig. 7. $I_{DS} - V_{DS}$ characteristics for the MoTe$_2$-MoS$_2$ TFET at several gate voltages ranging from $V_G = 0.35V$ to $V_G = 0.6V$, at $V_{BG-S} = 0.4V$ and $V_{BG-D} = 2.85V$.*

Although the results are not shown here, the same study was performed on the WTe$_2$-MoS$_2$ TFET and similar results were obtained: the influence of channel length on SS and ON current is the same, and the optimal backgate voltages at $V_{DS} = 0.3V$ for this device are found to be $V_{BG-S} = 0.85V$ and $V_{BG-D} = 0.65V$.

Under those conditions the WTe$_2$-MoS$_2$ TFET can yield SS below 5 mV/dec and ON currents beyond $2 \times 10^8$ µA/µm.

## Conclusion

By means of an atomistic tight-binding approach and self-consistent quantum simulations, we investigated several types of in-plane 2D material heterojunction based TFETs. Band alignment was highlighted as one of the most important parameters for TFET operation, and the influence of several design parameters on device performance was studied. Through careful selection of the materials system, channel length and backgate voltages, sub-threshold swings below 5 mV/dec and high $I_{ON}/I_{OFF}$ ratios ($> 10^8$) were reported at a low drain bias of 0.3V. Those in-plane heterojunction TFETs are therefore ideal candidates for ultra-low power operation.

## Acknowledgment

S.F. and E.K. were supported by the STC Center for Integrated Quantum Materials, NSF Grant No. DMR-1231319 and by ARO MURI Award W911NF-14-0247.

# Appendix A.

In this appendix, we describe the creation of the tight-binding Hamiltonians referred to in the body of the article.

For more in-depth information about the creation of the TB model itself and the way strain is handled, we refer the reader to the original article describing the model[24].

The details provided in this appendix are to be used in conjunction with the original article due to recurring notations and notions, and callbacks to parameters and calculations found in the article describing the TB model.

The 11×11 basis used to describe a single $MX_2$ unit cell in this TB model is the following

$$|d_{xz}\rangle, |d_{yz}\rangle, |p_x\rangle, |p_y\rangle, |p_z\rangle, |d_{xy}\rangle, |d_{x^2-y^2}\rangle, |d_{z^2}\rangle, |p_x\rangle, |p_y\rangle, |p_z\rangle$$

It contains the five *d* orbitals of the metal atom as well as the three *p* orbitals of each chalcogen atom, and is arranged as such due to symmetry considerations with respect to a *xy* mirror plane.

The original article describes the calculation of the 11×11 coupling Hamiltonians depending on the strain applied, and of the final, total Hamiltonian. However, as mentioned in Sec. 2, the use of a NEGF method requires a unit cell that can be reproduced along the transport direction, which is not the case of the basic $MX_2$ unit cell used in the TB model. We therefore use a unit cell that is twice the size of the original one, and will result in 22×22 Hamiltonians. In order to construct those Hamiltonians, we created matrices that describe the coupling between orbitals depending on the relative positions of the atoms they are associated with. There are nine "position pairs" leading to a coupling between orbitals and therefore nine such matrices, referred to as $\delta_1$ to $\delta_9$ and shown in Fig. 8.

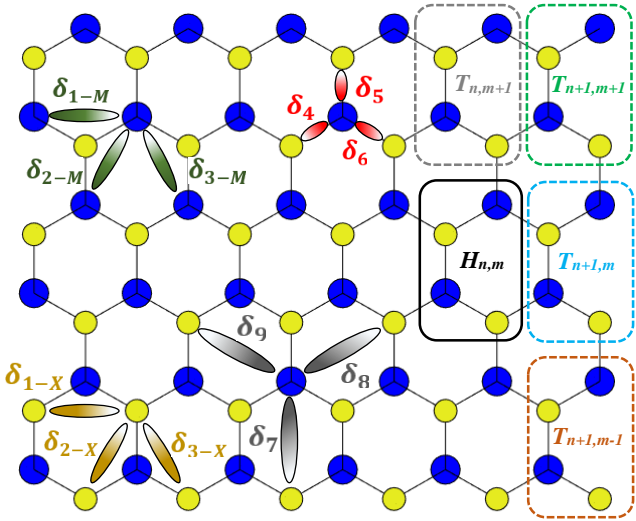

*Fig. 8. Lattice arrangement of a TMD. The highlighted $\delta_1$ through $\delta_9$ areas represent the coupling matrices between various orbitals on the considered atoms.*

In this figure, the direction of the coupling follows the color gradient (for instance, $\delta_4$ to $\delta_9$ represent the coupling from the X to the M atom).

The $\delta_i$ matrices are 11×11 matrices and are deduced from the Hamiltonians presented in the original article.

- $\delta_1$, $\delta_2$ and $\delta_3$ represent the coupling between orbitals located on atoms of the same type: either M-M or X-X coupling; they are split into separate $\delta_{i-X}$ and $\delta_{i-M}$ matrices to simplify Hamiltonian construction later on.
- $\delta_4$, $\delta_5$ and $\delta_6$ represent the first-neighbor coupling between orbitals located on atoms of a different type (X-M coupling).
- $\delta_7$, $\delta_8$ and $\delta_9$ represent the third-neighbor coupling between orbitals located on atoms of a different type (X-M coupling).

The $\delta_{1-X}$, $\delta_{1-M}$, $\delta_5$ and $\delta_7$ matrices are calculated from the Hamiltonians given in the original article as follows, where $H^{(n)}_{XY\,ij}$ is the $(i,j)$ element of the $H^{(n)}_{XY}$ matrix.

$$\delta_{1-X} = \begin{bmatrix} 0 & 0 & 0 & 0 & 0 & 0 & 0 & 0 & 0 & 0 & 0 \\ 0 & 0 & 0 & 0 & 0 & 0 & 0 & 0 & 0 & 0 & 0 \\ 0 & 0 & H^{(2)}_{BB\,11} & H^{(2)}_{BB\,21} & H^{(2)}_{BB\,31} & 0 & 0 & 0 & 0 & 0 & 0 \\ 0 & 0 & H^{(2)}_{BB\,12} & H^{(2)}_{BB\,22} & H^{(2)}_{BB\,32} & 0 & 0 & 0 & 0 & 0 & 0 \\ 0 & 0 & H^{(2)}_{BB\,13} & H^{(2)}_{BB\,23} & H^{(2)}_{BB\,33} & 0 & 0 & 0 & 0 & 0 & 0 \\ 0 & 0 & 0 & 0 & 0 & 0 & 0 & 0 & 0 & 0 & 0 \\ 0 & 0 & 0 & 0 & 0 & 0 & 0 & 0 & 0 & 0 & 0 \\ 0 & 0 & 0 & 0 & 0 & 0 & 0 & 0 & 0 & 0 & 0 \\ 0 & 0 & 0 & 0 & 0 & 0 & 0 & 0 & H^{(2)}_{DD\,11} & H^{(2)}_{DD\,21} & H^{(2)}_{DD\,31} \\ 0 & 0 & 0 & 0 & 0 & 0 & 0 & 0 & H^{(2)}_{DD\,12} & H^{(2)}_{DD\,22} & H^{(2)}_{DD\,32} \\ 0 & 0 & 0 & 0 & 0 & 0 & 0 & 0 & H^{(2)}_{DD\,13} & H^{(2)}_{DD\,23} & H^{(2)}_{DD\,33} \end{bmatrix}$$

$$\delta_{1-M} = \begin{bmatrix} H^{AA(2)}_{11} & H^{AA(2)}_{21} & 0 & 0 & 0 & 0 & 0 & 0 & 0 & 0 & 0 \\ H^{AA(2)}_{12} & H^{AA(2)}_{22} & 0 & 0 & 0 & 0 & 0 & 0 & 0 & 0 & 0 \\ 0 & 0 & 0 & 0 & 0 & 0 & 0 & 0 & 0 & 0 & 0 \\ 0 & 0 & 0 & 0 & 0 & 0 & 0 & 0 & 0 & 0 & 0 \\ 0 & 0 & 0 & 0 & 0 & 0 & 0 & 0 & 0 & 0 & 0 \\ 0 & 0 & 0 & 0 & 0 & H^{CC(2)}_{11} & H^{CC(2)}_{21} & H^{CC(2)}_{31} & 0 & 0 & 0 \\ 0 & 0 & 0 & 0 & 0 & H^{CC(2)}_{12} & H^{CC(2)}_{22} & H^{CC(2)}_{32} & 0 & 0 & 0 \\ 0 & 0 & 0 & 0 & 0 & H^{CC(2)}_{13} & H^{CC(2)}_{23} & H^{CC(2)}_{33} & 0 & 0 & 0 \\ 0 & 0 & 0 & 0 & 0 & 0 & 0 & 0 & 0 & 0 & 0 \\ 0 & 0 & 0 & 0 & 0 & 0 & 0 & 0 & 0 & 0 & 0 \\ 0 & 0 & 0 & 0 & 0 & 0 & 0 & 0 & 0 & 0 & 0 \end{bmatrix}$$

$$\delta_5 = \begin{bmatrix} 0 & 0 & 0 & 0 & 0 & 0 & 0 & 0 & 0 & 0 & 0 \\ 0 & 0 & 0 & 0 & 0 & 0 & 0 & 0 & 0 & 0 & 0 \\ H^{(1)}_{BA\,11} & 0 & 0 & 0 & 0 & 0 & 0 & 0 & 0 & 0 & 0 \\ 0 & H^{(1)}_{BA\,22} & 0 & 0 & 0 & 0 & 0 & 0 & 0 & 0 & 0 \\ 0 & H^{(1)}_{BA\,32} & 0 & 0 & 0 & 0 & 0 & 0 & 0 & 0 & 0 \\ 0 & 0 & 0 & 0 & 0 & 0 & 0 & 0 & 0 & 0 & 0 \\ 0 & 0 & 0 & 0 & 0 & 0 & 0 & 0 & 0 & 0 & 0 \\ 0 & 0 & 0 & 0 & 0 & 0 & 0 & 0 & 0 & 0 & 0 \\ 0 & 0 & 0 & 0 & H^{(1)}_{DC\,11} & 0 & 0 & 0 & 0 & 0 & 0 \\ 0 & 0 & 0 & 0 & H^{(1)}_{DC\,22} & 0 & H^{(1)}_{DC\,23} & 0 & 0 & 0 & 0 \\ 0 & 0 & 0 & 0 & H^{(1)}_{DC\,32} & 0 & H^{(1)}_{DC\,33} & 0 & 0 & 0 & 0 \end{bmatrix}$$

$$\delta_7 = \begin{bmatrix} 0 & 0 & 0 & 0 & 0 & 0 & 0 & 0 & 0 & 0 & 0 \\ 0 & 0 & 0 & 0 & 0 & 0 & 0 & 0 & 0 & 0 & 0 \\ 0 & 0 & 0 & 0 & 0 & 0 & 0 & 0 & 0 & 0 & 0 \\ 0 & 0 & 0 & 0 & 0 & 0 & 0 & 0 & 0 & 0 & 0 \\ 0 & 0 & 0 & 0 & 0 & 0 & 0 & 0 & 0 & 0 & 0 \\ 0 & 0 & 0 & 0 & 0 & 0 & 0 & 0 & 0 & 0 & 0 \\ 0 & 0 & 0 & 0 & 0 & 0 & 0 & 0 & 0 & 0 & 0 \\ 0 & 0 & 0 & 0 & 0 & 0 & 0 & 0 & 0 & 0 & 0 \\ 0 & 0 & 0 & 0 & H^{(3)}_{DC\,11} & 0 & 0 & 0 & 0 & 0 & 0 \\ 0 & 0 & 0 & 0 & 0 & H^{(3)}_{DC\,22} & H^{(3)}_{DC\,23} & 0 & 0 & 0 & 0 \\ 0 & 0 & 0 & 0 & 0 & H^{(3)}_{DC\,32} & H^{(3)}_{DC\,33} & 0 & 0 & 0 & 0 \end{bmatrix}$$

Due to the three-fold rotational symmetry of TMDs, all other $\delta_i$ matrices can be deduced from these four matrices via the transfer matrix $\hat{A}_R^{3\times 3}$, which describes a $\frac{2\pi}{3}$ counter-clockwise rotation.

$$\hat{A}_R^{3\times 3} = \begin{bmatrix} -1/2 & \sqrt{3}/2 & 0 \\ -\sqrt{3}/2 & -1/2 & 0 \\ 0 & 0 & 1 \end{bmatrix}$$

And can be transformed into a 11×11 matrix as such

$$\hat{U}_R = \begin{bmatrix} -1/2 & -\sqrt{3}/2 & 0 & 0 & 0 \\ -\sqrt{3}/2 & -1/2 & 0 & 0 & 0 \\ 0 & 0 & \hat{A}_R^{3\times 3} & 0 & 0 \\ 0 & 0 & 0 & \hat{A}_R^{3\times 3} & 0 \\ 0 & 0 & 0 & 0 & A_R^{3\times 3} \end{bmatrix}$$

Therefore, we can calculate some of the remaining $\delta_i$ matrices as follows

$$\delta_3 = \langle \hat{U}_R^\dagger | \delta_1 | \hat{U}_R \rangle, \delta_4 = \langle \hat{U}_R^\dagger | \delta_5 | \hat{U}_R \rangle, \delta_8 = \langle \hat{U}_R^\dagger | \delta_7 | \hat{U}_R \rangle$$

and, by applying the same rotation operation, we can obtain all remaining $\delta_i$ matrices

$$\delta_2^\dagger = \langle \hat{U}_R^\dagger | \delta_3 | \hat{U}_R \rangle, \delta_6 = \langle \hat{U}_R^\dagger | \delta_4 | \hat{U}_R \rangle, \delta_9 = \langle \hat{U}_R^\dagger | \delta_8 | \hat{U}_R \rangle$$

It is worth noting that by applying $\hat{U}_R$ on $\delta_3$, we obtain $\delta_2^\dagger$ and not $\delta_2$ because, due to the way the vectors were defined (see Fig. 8), $\vec{\delta_3}$ becomes $-\vec{\delta_2}$ via this $\frac{2\pi}{3}$ counter-clockwise rotation.

Now that all of the $\delta_i$ matrices have been calculated from parameters and Hamiltonians given in the original article, only the on-site energies remain to consider before we can construct the 22×22 Hamiltonians.

These on-site energies are the elements of the $H^{(0)}_{ii}$ matrices found in the original article, which we gather in a 11×11 diagonal matrix $E_i$, of which the diagonal element is

$$[H^{(0)}_{AA\,11}\ H^{(0)}_{AA\,22}\ H^{(0)}_{BB\,11}\ H^{(0)}_{BB\,22}\ H^{(0)}_{BB\,33}\ H^{(0)}_{CC\,11}\ \ldots$$
$$\ldots H^{(0)}_{CC\,22}\ H^{(0)}_{CC\,33}\ H^{(0)}_{DD\,11}\ H^{(0)}_{DD\,22}\ H^{(0)}_{DD\,33}]$$

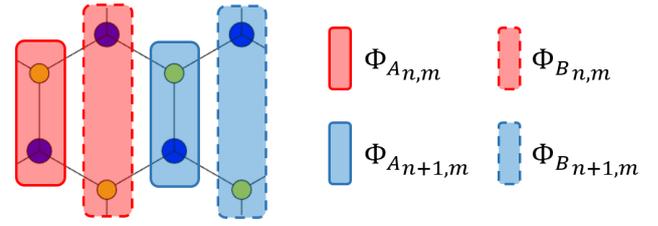

*Fig. 9. Visualisation of the A and B sublattices in two adjacent unit cells. Dashed (full) lines represent the A (B) sublattice, and their color indicates the unit cell to which they belong (red for the (n,m) unit cell, blue for the (n+1,m) unit cell)*

With all $\delta_i$ matrices and the on-site energies $E_i$ now calculated, the 22×22 Hamiltonians describing the $(MX_2)_2$ unit cell and its coupling with neighboring cells can be constructed.

Because the unit cell considered is twice the size of the original unit cell, two sublattices A and B (shown in Fig. 9) can be distinguished in each cell. Each of them is associated to an 11×11 basis

$$\Phi_A = |d_{xz}\rangle, |d_{yz}\rangle, |p_x\rangle, |p_y\rangle, |p_z\rangle, |d_{xy}\rangle, |d_{x^2-y^2}\rangle, |d_{z^2}\rangle, |p_x\rangle, |p_y\rangle, |p_z\rangle$$
$$\Phi_B = |d_{xz}\rangle, |d_{yz}\rangle, |p_x\rangle, |p_y\rangle, |p_z\rangle, |d_{xy}\rangle, |d_{x^2-y^2}\rangle, |d_{z^2}\rangle, |p_x\rangle, |p_y\rangle, |p_z\rangle$$

Each 22×22 Hamiltonian is therefore composed of four 11×11 matrices $\Psi_{X_{i,j}.Y_{k,l}}$. Those matrices are linear combinations of the $\delta_i$ and $E_i$ matrices, and describe the coupling between the X sub-lattice of the (i,j) unit cell and the Y sublattice of the (k,l) unit cell. They can be constructed by studying Fig.8 and selecting the appropriate couplings based on the atomic positions considered.

$H_{n,m}$ (the Hamiltonian of the unit cell) and $T_{i,j}$ (the Hamiltonians describing the coupling between the unit cell and adjacent cells), can be constructed as

$$H_{n,m} = \begin{pmatrix} \Psi_{A_{n,m}.A_{n,m}} & \Psi_{B_{n,m}.A_{n,m}} \\ \Psi_{A_{n,m}.B_{n,m}} & \Psi_{B_{n,m}.B_{n,m}} \end{pmatrix}$$

$$T_{i,j} = \begin{pmatrix} \Psi_{A_{n,m}.A_{i,j}} & \Psi_{B_{n,m}.A_{i,j}} \\ \Psi_{A_{n,m}.B_{i,j}} & \Psi_{B_{n,m}.B_{i,j}} \end{pmatrix}$$

with $i = [n-1, n, n+1]$ and $j = [m-1, m, m+1]$

Therefore, the final, 22×22 Hamiltonians describing each TMD are calculated as follows

$$H_{n,m} = \begin{pmatrix} E_i + \delta_5 + \delta_5^\dagger & \delta_6^\dagger + \delta_4 + \delta_{3-X} + \delta_{2-M}^\dagger \\ \delta_6 + \delta_4^\dagger + \delta_{3-X}^\dagger + \delta_{2-M} & E_i + \delta_7 + \delta_7^\dagger \end{pmatrix}$$

$$T_{n+1,m} = \begin{pmatrix} \delta_{1-X}^\dagger + \delta_{1-M}^\dagger + \delta_8^\dagger + \delta_9 & 0 \\ \delta_{3-M} + \delta_{2-X}^\dagger + \delta_6^\dagger + \delta_4 & \delta_{1-X}^\dagger + \delta_{1-M}^\dagger \end{pmatrix}$$

$$T_{n+1,m-1} = \begin{pmatrix} 0 & 0 \\ \delta_{3-X} & \delta_9 \end{pmatrix} \qquad T_{n+1,m+1} = \begin{pmatrix} 0 & 0 \\ \delta_{2-M}^\dagger & \delta_8^\dagger \end{pmatrix}$$

$$T_{n,m+1} = \begin{pmatrix} \delta_7 & \delta_{2-X}^\dagger \\ \delta_{3-M}^\dagger & \delta_5^\dagger \end{pmatrix} \qquad T_{n-k,m-l} = T_{n+k,m+l}^\dagger$$